\documentclass[prb,aps,superscriptaddress,reprint]{revtex4-1}

\usepackage{graphicx}

\usepackage{dcolumn}
\usepackage{bm}
\usepackage{setspace}
\usepackage{amsmath}
\usepackage{epsfig}

\newcommand{\fg}{f_{\mathrm{G}}}
\newcommand{\hg}{H_{\mathrm{loc}}}
\newcommand{\hu}{H_{\mathrm{uni}}}
\newcommand{\hip}{H_{\mathrm{IP}}}
\newcommand{\wg}{w_{\mathrm{loc}}}
\newcommand{\rmnp}{R_{\mathrm{P}}}
\newcommand{\msmnp}{M_{\mathrm{S,P}}}

\begin{document}

\title{Localized magnetic fields enhance the field-sensitivity of the gyrotropic  resonance frequency of a magnetic vortex}

\author{Jasper P.~Fried}
\author{Peter J.~Metaxas}
\email{peter.metaxas@uwa.edu.au}
\affiliation{School of Physics, M013, University of Western Australia, 35 Stirling Hwy, Crawley WA 6009, Australia.}%

\date{\today}



\begin{abstract}
We have carried out micromagnetic simulations of the gyrotropic resonance mode of a magnetic vortex in the presence of spatially localized and spatially uniform out-of-plane magnetic fields. We show that the field-induced change in the gyrotropic mode frequency is significantly larger when the  field is centrally localized over lengths which are comparable to or a few times larger than the vortex core radius. When aligned with the  core magnetization, such fields generate an additional confinement of the core. This confinement increases the vortex stiffness in the small displacement limit, leading to a resonance shift which is greater than that expected for a uniform out-of-plane field of the same amplitude. Fields generated by uniformly magnetized spherical particles having a fixed separation from the disk are found to generate analogous effects except that there is a maximum in the shift at intermediate particle sizes where field localization and stray field magnitude combine optimally to generate a maximum confinement.
\end{abstract}

\maketitle

\section{Introduction}

A magnetic vortex is a curled magnetization configuration with an out-of-plane magnetized nano-scale sized core \cite{Shinjo2000,Wachowiak2002}. Vortices arise spontaneously in (sub)micron-scale \cite{Guslienko2008} magnetic elements such as discs \cite{Cowburn1999} (as well as square \cite{Vogel2011a} or triangular \cite{Yakata2013} plates) and are of relevance for a range of applications ranging from radiofrequency signal generation \cite{Pribiag2007,Dussaux2010} and detection\cite{Jenkins2015} to cancer treatment \cite{Kim2010a}, data storage \cite{Yamada2007} and magnonics\cite{Huber2011,Han2013,Behncke2015}. Many  applications exploit the lowest frequency magnetic excitation of a vortex, the gyrotropic mode \cite{Guslienko2002,Choe2004,Ivanov2004,Guslienko2006a}, which corresponds to an orbit-like motion of the vortex core about the disk's center.  

An important characteristic of the gyrotropic mode is that its frequency, $\fg$, can be tuned by applying static out-of-plane\cite{Pribiag2007,deLoubens2009,Dussaux2010,Locatelli2011} or in-plane\cite{Yakata2013,Buchanan2006} magnetic fields. Uniform out-of-plane magnetic fields  modify both the element's magnetization configuration and the magnetostatic confinement of the core and, when sufficiently far below the saturation field of the disk, induce a change in $\fg$ which is a linear function of the field strength  \cite{Pribiag2007,deLoubens2009,Dussaux2010}. The combination of this field-linearity with the ability to electrically probe vortex dynamics (by fabricating a vortex-based spin torque nano-oscillator\cite{Pribiag2007,Dussaux2010,Kim2012a}, STNO), has application not only for field-tunable electronic oscillators \cite{Pribiag2007,Dussaux2010,Locatelli2011,Lebrun2014} 
but potentially also for intrinsically frequency-based sub-micron magnetic field sensors\cite{Braganca2010,Ryan2011}. The latter typically exploit the field dependence of the output frequency of a STNO  (e.g.~see \cite{Braganca2010,Ryan2011,Srimani2015} for non-vortex based devices) for frequency-based\cite{Braganca2010,Mizushima2010,Inoue2011,Atalay2015,Metaxas2015} field sensing. One potential application of such a sensor is in the development of frequency-based nano-scale devices to detect (biofunctionalized) magnetic nanoparticles\cite{Metaxas2015} (MNPs) for in-vitro bio-sensing and point-of-care medical diagnostics  \cite{Gaster2009}.

In this work we show that central, localized out-of-plane fields (such as those generated by MNPs) produce shifts in the gyrotropic frequency greater than those induced by uniform out-of-plane fields having the same amplitude. It is shown that this is due to an increase in the vortex stiffness as a result of the out-of-plane magnetization of the core preferentially aligning with the strongest part of the localized field. For  the particular case of MNPs whose lower surfaces are separated from the disk by a fixed distance, we demonstrate that the frequency shift is characterized by a clear maximum at intermediate particle sizes,  a result of an optimized combination of the amplitude and localization of the stray field created by the MNP.   We also note that short range structural disorder (as well as domain-generated stray fields\cite{Heldt2014}) can also pin vortices\cite{Uhlig2005} (with a potential for quantum depinning\cite{Zarzuela2012,Zarzuela2013}) and that these defects have also been shown to be capable of modifying the gyrotropic mode frequency\cite{Silva2008a,Vansteenkiste2009,Compton2010,Dussaux2010,Min2011,Chen2012b}. 

\section{Micromagnetic simulation}

Our results were obtained using finite difference micromagnetic simulations of the gyrotropic mode using MuMax3 \cite{Vansteenkiste2014}.
  We will focus on simulation results for NiFe discs with radius $R=192$ nm, thickness $L=30$ nm, saturation magnetization $M_{\mathrm{S}}=800$ kA/m, exchange stiffness $A_{\mathrm{ex}}=13$ pJ/m, magnetic damping $\alpha=0.008$,  no intrinsic anisotropy and a cell size of $2\times2\times3.75$ nm$^3$. To begin with, the system is initialized with a vortex-like  magnetic configuration and  allowed to relax using MuMax3's internal relaxation routine which time evolves the magnetization (without precession) using energy- and then torque-minimization as stopping criteria\cite{Vansteenkiste2014}.  A transverse magnetic field sinc pulse is then applied with an amplitude of 2 mT, a time offset  of 300 ps and a cut-off frequency of 30 GHz. This induces a displacement of the vortex core (as well as some higher frequency excitations\cite{Ivanov2002,Ivanov2005,Park2005,Zaspel2005,Buchanan2006,Aliev2009}) which is followed by a damped gyrotropic motion of the core around the disk's center. Fourier analysis of the $x$-component of the spatially averaged magnetization is  used to extract $\fg$.  Good agreement between  MuMax3,  OOMMF \cite{oommf} and FinMag (derived from Nmag \cite{Fischbacher2007}) was confirmed for a number of test cases\footnote{Simulations for an identically sized disk ($R=192$ nm, $t=30$ nm) with a $3\times 3\times 3$ nm$^3$ cell size were run in OOMMF for zero field (agreement with MuMax within 1.1\%) and in the presence of a localized $\hg$ field (Sec.~\ref{sgaussian}) with $\wg=100$ nm (agreement within 1\%). There was also good agreement in the bare disk frequency using the eigenmode solver in FinMag for a $R=96$ nm disk however a larger meshing had to be used at the disk edges due to memory constraints.}.

\begin{figure}[h]
	\centering
	\includegraphics[width=8.5cm]{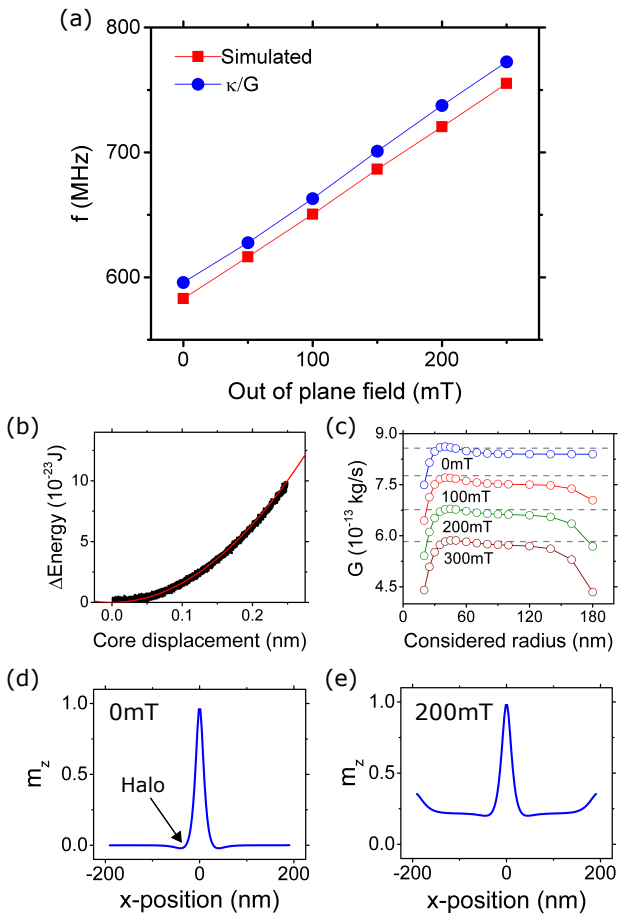}
	\caption{(a) Simulated and calculated gyrotropic mode frequencies, $\fg$ [the latter calculated using Eq.~(\ref{ef})],  for spatially uniform out-of-plane fields. (b) Plot of the change in energy (relative to the energy at zero core displacement) as a function of core displacement as measured during field-pulse-induced gyrotropic motion  in zero out-of-plane field. The first 25 ns of motion are disregarded due to the high frequency spin waves produced in this period. A quadratic fit to the energy is shown by the red curve. (c) Gyroconstant calculated considering different radii from the disk center for four different uniform out-of-plane fields. The gray dashed lines represent the value of the gyroconstant calculated using\cite{deLoubens2009} $G(H)=G(0)(1-\cos\theta)$ where $\cos\theta$ is taken at the magnetostatic halo surrounding the vortex core. (d) Out-of-plane magnetization for a centered (non-displaced) vortex along a slice through the disk's center in zero out-of-plane field and (e) under a spatially uniform 200 mT out-of-plane field.}
	\label{foop}
\end{figure}

\section{Uniform out of plane magnetic fields}

We will first consider the case of a vortex in a spatially uniform out-of-plane field \cite{deLoubens2009}, $\hu$. Fig.~\ref{foop}(a) shows the simulated gyrotropic frequency under various amplitude out-of-plane fields aligned with the vortex core. As previously observed, the frequency varies linearly with $\hu$. We were able to quantitatively reproduce the simulated frequencies to within 2\% [blue circles in Fig.~\ref{foop}(a)] using \cite{Guslienko2002}
\begin{equation}
2\pi\fg=\frac{\kappa}{G}
\label{ef}
\end{equation}
where $\kappa$ is the vortex stiffness coefficient and $G$ is the gyroconstant. The remainder of this section is focused on the extraction of $\kappa$ and $G$ to obtain the calculated $\fg=\kappa/(2\pi G)$ values shown in Fig.~\ref{foop}(a).

$\kappa$, the stiffness coefficient describes the harmonic scaling of the vortex energy, $W$, with lateral in-plane displacement, $X$, measured radially from the disk's center: $W(X)=W(0)+\frac{1}{2}\kappa X^2 +\mathcal{O}(X^4)$. For uniform or null out-of-plane fields, this confinement is dominated by  dipolar effects\cite{Guslienko2006a} however dynamic exchange fields are also present. For each simulation, we extracted $\kappa$ from a parabolic fit to the total energy of the system plotted against the dynamic displacement of the core as measured during the field-pulse-induced gyrotropic motion [e.g.~Fig.~\ref{foop}(b)]. As shown previously\cite{Buchanan2006}, this  dynamic approach, which analyzes the energy of the moving vortex core, produces a more accurate prediction of the gyrotropic frequency than a static method in which the total energy is calculated for displaced cores at equilibrium which have been shifted by static in-plane fields. This said, the static method will be  instructive in visualizing the influence of localized fields on the vortex stiffness. 

The gyroconstant, $G$, determines the magnitude of the gyrovector, $\mathbf{G}=G\hat{\mathbf e}_Z$, in the Thiele equation describing vortex dynamics \cite{Thiele1973,Huber1982,Guslienko2002,Ivanov2007}.  The gyroconstant can be calculated from the vortex spin structure using
\begin{equation}
G=\frac{M_s L}{\gamma}\iint_{A}\mathbf{m}\cdot \left (\frac{d\mathbf{m}}{dx}\times\frac{d\mathbf{m}}{dy}\right )dx dy
\label{eg}
\end{equation}
where $\gamma$ is the gyromagnetic ratio and $\mathbf{m}$ is the unit-length magnetization vector. Given that $\mathbf{G}$ acts along the z-axis this equation can be shown to be equal to the often quoted \cite{Guslienko2002,Huber1982} $\mathbf{G} = M_S L/\gamma\iint \sin(\theta)(\mathbf{\nabla} \theta \times \mathbf{\nabla} \phi)\,dx\,dy$ where $\theta$ and $\phi$ are the polar and azimuthal angles of the magnetization respectively. Theoretically, for a zero out-of-plane field Eq.~(\ref{eg}) yields $G=2\pi qpLM_s/\gamma$ where $q=1$ is the vorticity and $p=\pm 1$ is the core polarity aligned along $\pm \hat{\mathbf e}_Z$.  

We numerically calculated the gyroconstant associated with the static, non-displaced core for the studied $\hu$ values using Eq.~(\ref{eg}) by integrating a thickness-averaged $\mathbf{m}$ \footnote{We observed a 0.4\% change in G across different layers of the disk.} over the entire disk area. However, inserting this calculated value of $G$ into Eq.~(\ref{ef}) led to a value for $\fg$ which was  significantly lower than the simulated value. For example, for a uniform  out-of-plane field of 200 mT the gyroconstant calculated using this method is $\approx 36\%$ smaller than what is expected taking the simulated frequency and extracted $\kappa$ and solving for $G$ (i.e.~$5.17\times10^{-13}$ kgs$^{-1}$ compared to $7.03\times10^{-13}$ kgs$^{-1}=\kappa/2 \pi f_g$). To attempt to  understand this discrepancy, we  calculated $G$ by considering only the spin structure within a given radius of the disk center for four different $\hu$ values. The resultant data is shown in Fig.~\ref{foop}(c). $G$ reaches a clear maximum when integrating over a radius close to the edge of the vortex core ($\approx 40$ nm). Notably, at this point, the value of $G$ closely corresponds to the gyroconstant expected from the simulated frequency and extracted stiffness coefficient. This peak is present for all out-of-plane field amplitudes and is due to the `magnetostatic halo' [see Fig.~\ref{foop}(d)] surrounding the vortex core as a result of its demagnetization field \cite{Gaididei2010}. The drop-off in $G$ at large radii for $\hu >0$ is due to the out-of-plane canting of spins near the disk edge [e.g.~Fig.~\ref{foop}(e)].  Encouragingly, the $G$ calculated using this method corresponds closely to the value predicted by the equation $G(H)=G(0)(1-\cos\theta)$ given by de Loubens \textit{et al} \cite{deLoubens2009} for uniform out-of-plane fields where $\cos\theta$ is given as the polar angle of the magnetization at the disk's edge. The value of $G$ calculated using the above expression for $G(H)$ is shown by the gray dashed lines in Fig.~\ref{foop}(c) however $\cos\theta$ has been extracted at the center of the magnetostatic halo. This result suggests that although there is a divergence of the magnetization far away from the vortex core, it is the local spin structure around the core which is relevant for the  small amplitude oscillations considered here ($<1$ nm).

In Fig.~\ref{foop}(c) we see that $G$ reduces considerably when $\hu$ is increased and it is this reduction in $G$ which is primarily responsible for the linear increase in $\fg$ with increasing $\hu$. In fact, $\kappa$ is reduced at large $\hu$ values where it becomes easier to shift the vortex laterally. Below, localized out-of-plane fields will also be shown to increase $\fg$ however that increase will be demonstrated to be primarily due to a \textit{increase} in $\kappa$ induced by the localized field.

\begin{figure}[h]
	\centering
	\includegraphics[width=8.5cm]{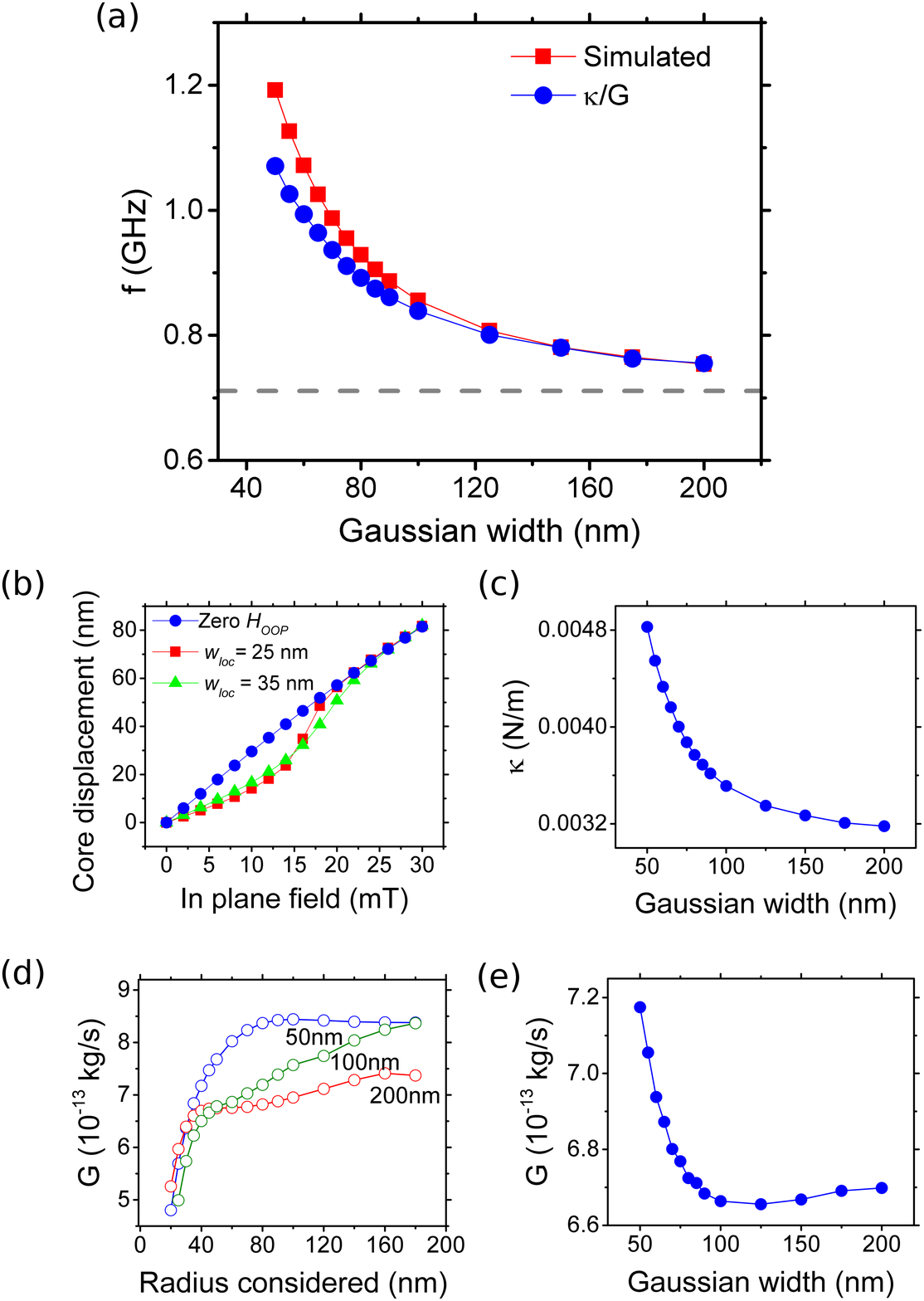}
	\caption{(a) Simulated and calculated gyrotropic mode frequencies, $\fg$ [the latter calculated using Eq.~(\ref{ef})], in the presence of 200 mT amplitude Gaussian [Eq.~(\ref{eg})] out-of-plane fields of various widths (the width is approximately  1.2 times the half-width-half-maximum (HWHM) of the Gaussian field profile). The simulated frequency for a spatially uniform 200 mT out-of-plane field is shown by the gray dashed line. (b) Vortex core displacement as a function of static in-plane field for two Gaussian fields and zero out-of-plane field. (c) Stiffness coefficient versus  Gaussian field width. (d) Gyroconstant calculated considering different radii from the disk center for three different Gaussian field widths. (e) Gyroconstant dependence on Gaussian field width where $G$ is calculated considering only the vortex structure within 40 nm of the disk's center.}
	\label{fgaussian}
\end{figure} 

\section{Effect of spatially localized fields}

\subsection{Gaussian fields}\label{sgaussian}

To study the effect of spatially localized out-of-plane fields on the gyrotropic frequency, we first consider centralized fields with a two-dimensional (2D) Gaussian-like profile: 
\begin{equation}
\hg=A_{\mathrm{0}}\exp\left[-\left(\frac{r}{\wg}\right)^2\right].
\label{eg}
\end{equation}
Here $\wg$ is a width parameter approximately equal to 1.2 times the half-width-half-maximum (HWHM) of the field profile and $r$ is the lateral distance from the center of the disk. $\mu_0 A_0$ is fixed at +200 mT making $\hg$ aligned with the vortex polarity ($p=+1$). As shown in Fig.~\ref{fgaussian}(a) localized Gaussian fields  significantly increase $\fg$ as compared to the action of a spatially uniform 200 mT out-of-plane field [gray dotted line in Fig.~\ref{fgaussian}(a)]. Furthermore, this frequency enhancement becomes larger as the Gaussian field becomes more localized.

To begin to understand the above frequency behavior (and confirm its link it to a $\hg$-induced confinement), we first look at how the vortex core moves laterally in response to static, uniform in-plane fields, $\hip$, in the  presence of a centrally localized field, $\hg$. By doing this we can explicitly probe the (static) confinement of the core across the disc. In the complete absence of an out-of-plane field, the core position varies linearly with $\hip$ at low fields (low displacements) [Fig.~\ref{fgaussian}(b)]. For higher displacements the response to field is slightly weaker,  consistent with an increased (anharmonic \cite{Sukhostavets2013}) confinement when the core moves closer to the disk's edge under the action of $\hip$. If we add a localized field  however, the $\hip$-induced core displacement is clearly  lower, but only if the displacement is comparable or smaller than the HWHM of the localized field. Indeed, at larger displacements, the response to $\hip$ is similar for both localized and uniform out-of-plane fields. This result explicitly confirms the $\hg$-induced confinement (or stiffening) of the vortex core which arises because keeping the core within the central region minimizes the Zeeman energy associated with the interaction between $\hg$ and the vortex core magnetization.   

\begin{figure}[h]
	\centering
	\includegraphics[width=7cm]{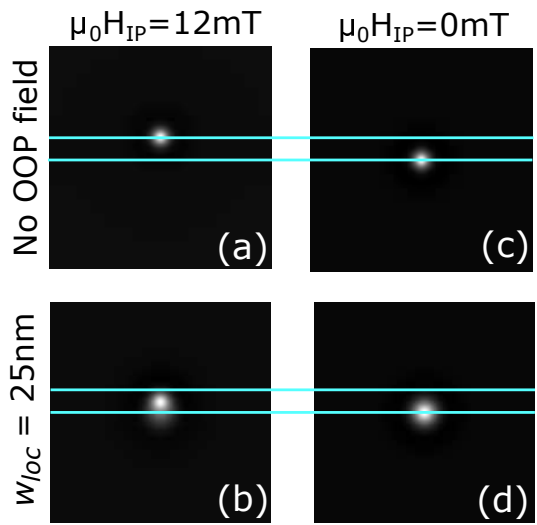}
	\caption{Visualizations of the out-of-plane component of the disk magnetization (white) for: (a) $\mu_0\hu =0$ mT and $\mu_0\hip$=12 mT, (b) a $\hg$ with $\wg$=25 nm and $\mu_0\hip$=12 mT, (c) $\mu_0\hu =0$ mT and $\mu_0\hip$=0 mT and (d) a $\hg$ with  $\wg$=25 nm and $\mu_0\hip$=0 mT. The light blue lines reference the centered and $\hip$-displaced core positions in zero out-of-plane field. In (a) and (c) the white part of the image corresponds to the core. In (b) and (d) the out-of-plane core magnetization is convoluted with the $\hg$-induced out-of-plane canting in the disk's center (which also translates to white coloring).}
	\label{fshifted}
\end{figure}

The influence of $\hg$ on the $\hip$-induced shift is visualized directly in Fig.~\ref{fshifted} where we compare the equilibrium static positions of the vortex core (white) for $\hip=12$ mT in two cases: a vortex with no out-of-plane field [Fig.~\ref{fshifted}(a)] and a vortex subject to a $\hg$ field with $\wg=25$ nm [Fig.~\ref{fshifted}(b)]. The core has clearly been displaced a smaller distance for the case of a localized field, with the core remaining within the strong part of the $\hg$ profile (visible as a broad out-of-plane magnetization component at the disk's center). Reference images of the unshifted vortex core at $\hip=0$ mT are given in Figs.~\ref{fshifted}(c,d). 

\begin{figure}[h]
	\centering
	\includegraphics[width=8.5cm]{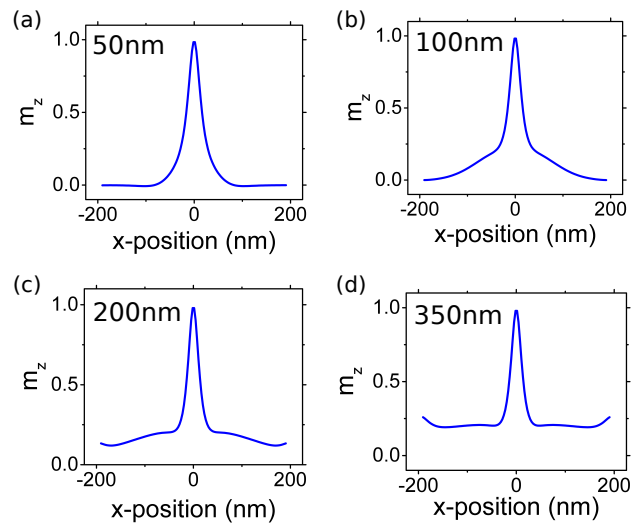}
	\caption{Plots of the out-of-plane magnetization for a slice through the disk center for a vortex in Gaussian fields of width, $\wg$ of (a) 50 nm (b) 100 nm (c) 200 nm and (d) 350 nm.}
	\label{fmz}
\end{figure}

By looking at Fig.~\ref{fshifted}(d) and Fig.~\ref{fmz}, the latter showing profiles of the out-of-plane component of the magnetization across the disk, one also sees that the presence of $\hg$ clearly modifies the magnetic configuration within the disk. For small $\wg$ the equilibrium core structure itself is changed [Fig.~\ref{fmz}(a)]: the magnetostatic halo is less sharp and the core widens. Intermediate $\wg$ values [Figs.~\ref{fmz}(b,c)] generate a clear out of plane magnetized region around the core while large $\wg$ values [Fig.~\ref{fmz}(d)], which lead to a broad field profile, result in a magnetization profile which is similar to that seen for uniform out-of-plane fields [Fig.~\ref{foop}(e)]. For small displacements of the core in all of these cases however, the confining potential nevertheless remains  close to harmonic  and we have again used the  method previously described to calculate (dynamic) values of $\kappa$. Consistent with the results seen in Figs.~\ref{fgaussian}(a,b) and \ref{fshifted}, $\kappa$ increases strongly for narrower localizations of the Gaussian field [as shown in Fig.~\ref{fgaussian}(c)], again confirming the $\hg$-induced confinement. 

We also calculated the gyroconstant in the presence of the $\hg$ fields, again considering different radii as done previously. The resultant data are shown in Fig.~\ref{fgaussian}(d) for three Gaussian field profiles. Notably, the peak in $G$  which was clearly visible in Fig.~\ref{foop}(c) disappears (or for $\wg=200$ nm becomes much less prominent) due to the absence of a deep magnetostatic halo for the studied $\wg$ values [Fig.~\ref{fmz}]. For the two broader Gaussians, $G$ increases with the considered radius since the non-uniform $\hg$ induces a canting which depends on the distance from the center of the disk [as seen in Figs.~\ref{fmz}(b,c)]. For the narrowest Gaussian, $G$  becomes flat at large considered radii. This is because the narrow localization of $\hg$ leads to a quasi-null canting of $\mathbf{m}$ away from the center. However, $G$ grows quickly at small and intermediate considered radii,  due to the $\hg$-induced broadening of the core [Fig.~\ref{fmz}(a)]. In  Fig.~\ref{fgaussian}(e) we show the extracted $G$ values versus $\wg$ where $G$ was again calculated using a considered radius of  40 nm (i.e.~analyzing the magnetization in the core's immediate vicinity). Like $\kappa$, $G$ also increases at small $\wg$ but to a lesser extent.

The extracted $G$ and dynamic $\kappa$   yield reasonable quantitative agreement between the simulated frequencies and the frequency predicted by Eq.~(\ref{ef}) [red squares in Fig.~\ref{fgaussian}(a)].  This said, the agreement is clearly best at large $\wg$. This is perhaps not unexpected however since wider profiles result in  weaker deformation of the magnetization near the core. 
We also emphasize that, in contrast to the case of a uniform out-of-plane field the growth in $\fg$ at small $\wg$ [Fig.~\ref{fgaussian}(a)] is driven by increased confinement [i.e.~$\kappa$ in Fig.~\ref{fgaussian}(c)] rather than by changes in  $G$ [Fig.~\ref{fgaussian}(e)].

\begin{figure}[htbp]
	\centering
	\includegraphics[width=8.5cm]{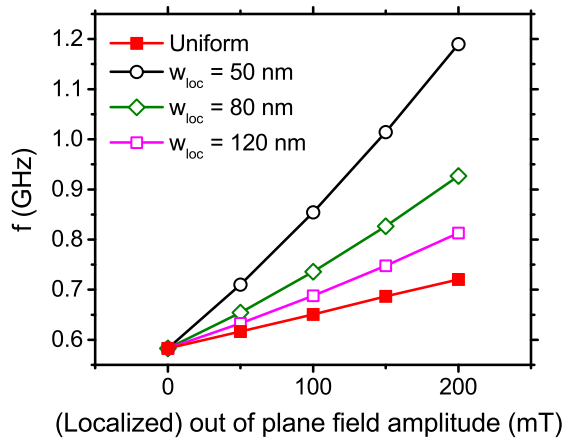}
	\caption{Simulated gyrotropic mode frequency plotted against the amplitude of localized, Gaussian fields with different widths. Data for a spatially uniform field [taken from Fig.~\ref{foop}(a)] is shown for comparison.}
	\label{famp}
\end{figure}

Up until now we have considered only the modifications to $\fg$ induced by changing $\wg$. 
Fig.~\ref{famp} however shows data analogous to that  in Fig.~\ref{foop}(a), demonstrating the change in $\fg$ induced when modifying the \textit{amplitude} of the localized fields. The change to $\fg$ per unit of field amplitude, which can be thought of as a `field sensitivity', is notably more than five times larger for a localized Gaussian field  with $\wg = 50$ nm than that observed for a spatially uniform field. Consistent with the results shown in Fig.~\ref{fgaussian}(a), this sensitivity enhancement reduces as the field profile is made broader (i.e.~when $\wg$ increases).

\begin{figure*}[htbp]
	\centering
	\includegraphics[width=15cm]{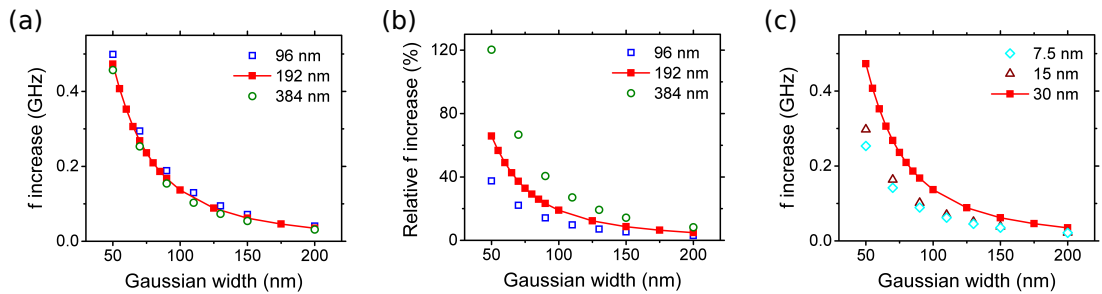}
	\caption{The (a) absolute and (b) relative change in the gyrotropic frequency, $\fg$ (compared to its value in $\hu=200$ mT) as a function of Gaussian width for disks of radii $R=$ 96, 192 and 384 nm and a constant thickness of $L=$ 30 nm. (c) The absolute change in $\fg$ for disks of thickness $L=$ 7.5, 15 and 30 nm and a constant radius of $R=$ 192 nm.}
	\label{fsizes}
\end{figure*}

Finally for this section, we demonstrate that the trends observed in Fig.~\ref{fgaussian}(a) apply to other disk geometries\footnote{Other disk geometries were simulated using a cell size of $3\times3\times3.75$ nm$^3$.}. Fig.~\ref{fsizes}(a) shows the frequency increase (relative to $\fg$ under $\hu = 200$ mT)  as a function of $\wg$ for three different disk radii. The increase in $\fg$ is found to be almost independent of the disk radius. This demonstrates for these radii that the frequency increase (at least for small oscillations in the presence of this strong 200 mT localized field) is largely independent of the `intrinsic' core confinement which is defined by the disc geometry. The increase instead results from the interaction between the localized field and the vortex core, the size of the latter being independent of the lateral disk dimensions \cite{Usov1993}. Along these same lines, since $\fg$ is intrinsically smaller at larger disk radii\cite{Guslienko2002,Park2005,Guslienko2006a,deLoubens2009} due to a weaker core confinement, the \textit{relative} frequency shift induced by a localized field (i.e.~as a percentage) will be higher for  wider disks [see Fig.~\ref{fsizes}(b)].

Analogous behavior to that shown in Fig.~\ref{fgaussian}(a) was also seen when reducing the disk thickness. The absolute change in $\fg$ did however reduce as the disk became thinner   [Fig.~\ref{fsizes}(c)]. This  reduction was found to be driven by a reduced dynamic $\kappa$ (confirmed at $\wg=90$ nm), indicating a lower Gaussian-field-induced core confinement at smaller thicknesses. This reduced confinement is perhaps not surprising as decreasing the disk thickness leads to a smaller vortex core volume (the core is narrower \cite{Usov1993} and its height reduces), lowering the $\hg$-associated Zeeman energy which drives the  confinement.

\subsection{Dipole fields and width-dependent Gaussian fields}

\begin{figure}[htbp]
	\centering
	\includegraphics[width=8.3cm]{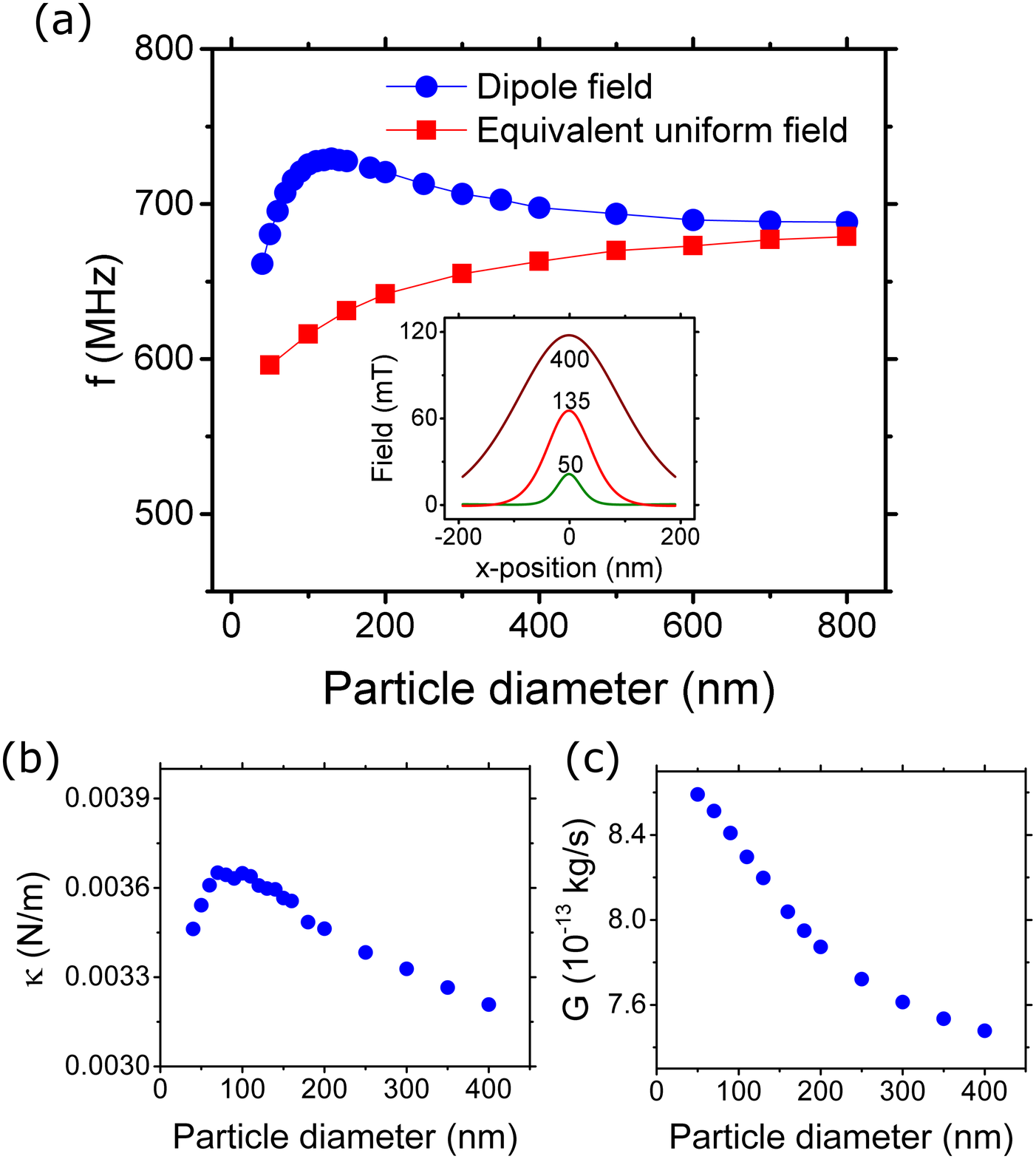}
	\caption{(a) Simulated frequency as a function of MNP diameter ($2\rmnp$). The red squares shows the simulated frequency in a uniform out-of-plane field of the same amplitude as the stray field created by the MNP as calculated at the disk center. The inset shows the out-of-plane component of the dipole field for three particle diameters below, above and near the observed peak in $\fg$. (b) Stiffness coefficient as a function of particle diameter. (c) Gyrovector calculated by considering the spin structure within 40 nm of the disk center as a function of particle diameter.
	}
	\label{fmnp}
\end{figure}

Localized magnetic fields can also be generated by uniformly magnetized, spherical MNPs. The in-plane components of the field generated by a submicron particle has previously been used to shift the vortex core position and probe the anharmonicity of the disk's intrinsic confining potential\cite{Sukhostavets2013}. Here we consider the case of a field generated by a \textit{centralized} MNP with radius $\rmnp$  whose lower surface is at a fixed distance (10 nm) from the upper surface of the disk (i.e.~the height of the center of the MNP from the top of the disk will be $\rmnp+10$ nm). To minimize simulation time the field created by a MNP with saturation magnetization $\msmnp$ has been modelled as that of a dipole with moment $\frac{4}{3} \pi \rmnp^3\msmnp$ where $\msmnp=200$ kAm$^{-1}$. 
To confirm the validity of this simplification, we determined the gyrotropic frequency observed when a solid ferromagnetic sphere of diameter 100 nm ($A_{\mathrm{ex}}=13$ pJ/m) was placed above the center of the disk (in this case we used a smaller disk with $R=96$ nm to minimize simulation time and memory use) and explicitly simulated it together with the disk in a 200 mT uniform out-of-plane field.   When compared to the simulated $\fg$ in an equivalent dipole field there was a relatively small discrepancy of $\approx 2.6\%$ which was further reduced to 0.3\% when adding a strong $z$-axis oriented anisotropy to the sphere ($K_1=10^7$ J/m$^3$). The latter tends to fix the sphere's magnetization in the out-of-plane orientation, suggesting that the worse agreement for the zero-anisotropy sphere was due to changes in the sphere's magnetic configuration induced by the magnetic stray field of the disk.  We also note that the simulations below were performed with no external out-of-plane field. However, a field would usually have to be applied experimentally in the case of a superparamagnetic particle and indeed we found analogous results to those given below [$\fg$ versus $\rmnp$ in Fig.~\ref{fmnp}(a)] for simulations in a 200 mT external out-of-plane field (sufficient to induce the aforementioned $\msmnp$ value\cite{Metaxas2015}).

In Fig.~\ref{fmnp}(a), we show results obtained for dipoles equivalent to $+z$-magnetized MNPs of various sizes positioned above the disk as detailed above.  In contrast to the case of the Gaussian field, $\fg$ displays a maximum at some intermediate particle size. This behavior cannot be explained simply by the stray field changing in magnitude as $\rmnp$ changes. Indeed, if we run simulations to determine $\fg$ in the presence of uniform out-of-plane fields equivalent in magnitude to the MNP field (as calculated in the center of the disk) we see a monotonically increasing $\fg$ [red squares in Fig.~\ref{fmnp}(a)] with no peak. This latter growth in $\fg$ is consistent with the bigger particles generating bigger fields and thus bigger changes in $\fg$ [i.e.~as per Fig.~\ref{foop}(a)]. Instead, the peak in $\fg$ observed for the dipole fields at intermediate $\rmnp$ can be correlated with a  maximum in the confinement of the core, manifested as a peak in $\kappa$ [Fig.~\ref{fmnp}(b)]. Note however that the maximum $\fg$ occurs at a slightly higher diameter than that which leads to the maximum $\kappa$ due to the diameter dependence of $G$ which increases [thus reducing $\fg$, as per Eq.~(\ref{ef})] as the particle diameter becomes small   [Fig.~\ref{fmnp}(c)]. 

In the inset of Fig.~\ref{fmnp}(a) we show the out-of-plane component of the dipolar MNP field for particle diameters located around the point at which the peak in $\fg$ lies.  The peak in confinement  can be understood as follows. At small particle diameters (e.g.~$2\rmnp =50$ nm), the field is highly localized however the magnetic moment of the MNP (and thus the stray field amplitude) is low. This results in a weak confinement. A weak confinement also occurs for large particles (e.g.~$2\rmnp =400$ nm) which generate strong but nevertheless broad (and thus weakly localized)  spatial field profiles.  Between these two extrema however is  some intermediate particle size (e.g.~$2\rmnp = 135$ nm) where there is an optimal combination of field strength and localization which maximizes the core confinement, thus leading to a large $\fg$.

\begin{figure}[htbp]
	\centering
	\includegraphics[width=8.5cm]{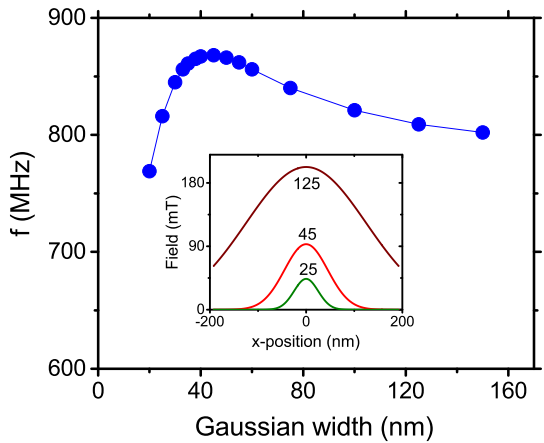}
	\caption{Simulated frequency in the presence of a Gaussian field whose amplitude has been modified to depend upon its width and thus generate analogous data to that in Fig.~\ref{fmnp}(a). Inset shows a central slice through the field profile for three different Gaussian widths.}
	\label{fmodgauss}
\end{figure}

We can also reproduce the above tendencies using Gaussian fields which have been  scaled by a factor $\propto\wg^3/(\wg+d)^3$. This scaling mimics some characteristics of the $\rmnp$-dependent dipole field. $d$ is chosen to be 25 nm as this is the distance used in our simulations between the bottom of the particle and center of the disk. The numerator in the above scaling factor describes the magnetic moment dependence on particle radius (here equivalent to $\wg$) whereas the denominator describes the field behavior as the dipole center moves further away from the disk center due to an increasing particle radius. The simulated $\fg$ values  are shown in Fig.~\ref{fmodgauss}. Unlike the results for Gaussian fields of constant amplitude, $\fg$ now exhibits a peak at intermediate $\wg$ values, analogous to the frequency behavior seen for MNPs when changing $\rmnp$ [Fig.~\ref{fmnp}(a)]. The corresponding one dimensional Gaussian field profiles are shown in the inset of Fig.~\ref{fmodgauss} for three different $\wg$ values. Again, as we increase $\wg$ we see a transition from a weak, highly localized field to a strong, broadly localized field with maximum confinement occurring at an intermediate $\wg$. 

\section{Conclusion}

We have shown that the sensitivity of the vortex gyrotropic mode frequency to out-of-plane fields can depend strongly on the spatial localization of those fields. Centralized, out-of-plane magnetic fields localized over lengths which are comparable to or a few times larger than the vortex core radius induce significantly larger changes in the vortex gyrotropic frequency than that generated by a spatially uniform out-of-plane field of the same amplitude. This behavior is consistent with an increase in the vortex stiffness as a result of the out-of-plane magnetization of the core preferentially aligning with the strongest part of the localized  field which generates an additional vortex core confinement. In the case of fields which approximate those generated by magnetic particles of varying radius, the frequency was observed to be maximized for some intermediate particle size which led to an optimized combination of field amplitude and field localization. This  may be relevant for vortex-based MNP sensors exploiting changes of the gyrotropic frequency induced by localized MNP fields\cite{Wohlhuter2015}. We finally note that this work has focused on very low amplitude excitations and highlight the fact that gyrotropic motion outside of the strongest part of the localized field will result in weaker changes  to $\fg$  (as observed by Min \textit{et al}.\cite{Min2011}~for large amplitude oscillations around pinning sites generated by changes in saturation magnetization).

\begin{acknowledgements}

This work was supported by the Australian Research Council's Discovery Early Career Researcher Award scheme (DE120100155), a research grant from the United States Air Force (Asian Office of Aerospace Research and Development, AOARD), the University of Western Australia (including the ECRFS, RCA and RDA schemes), NeCTAR (National eResearch Collaboration Tools and Resources, supported by the Australian Government through the National Collaborative Research Infrastructure Strategy) and  by resources provided by the Pawsey Supercomputing Centre with funding from the Australian Government and the Government of Western Australia.  The authors thank Joo-Von Kim, Manu Sushruth, Mikhail Kostylev, Vincent Cros, Rebecca Carey, Maximilian Albert and Hans Fangohr for useful discussions, advice and/or assistance. 

\end{acknowledgements}

\end{document}